\documentstyle[12pt]{article}

\begin{document}

\topmargin 0pt \oddsidemargin 5mm

\setcounter{page}{1} 
\vspace{2cm}

\begin{center}
{\bf Data processing in quantum information theory }\\ \vspace{5mm} {\large %
A.E. Allahverdyan, D.B. Saakian }\\ \vspace{5mm} {\em Yerevan Physics
Institute}\\ {Alikhanian Brothers St.2, Yerevan 375036, Armenia }\\
\end{center}

\vspace{5mm} \centerline{\bf{Abstract}} The strengthened data processing
inequality have been proved. The general theory have been illustrated on the
simple example.



\vspace{5mm} Quantum information theory is a new field with potential
applications for the conceptual foundation of quantum mechanics. It appears
to be the basis for a proper understanding of the emerging fields of quantum
computation, communication and cryptography [1-4]. Quantum information
theory concerned with quantum bits (qubits) rather than bits. Qubits can
exist in superposition or entanglement states with other qubits, a notion
completely inaccessible for classical mechanics. More general, quantum
information theory contains two distinct types of problem. The first type
describes transmission of classical information through a quantum channel
(the channel can be noisy or noiseless). In such scheme bits encoded as some
quantum states and only this states or its tensor products are transmitted.
In the second case arbitrary superposition of this states or entanglement
states are transmitted. In the first case the problems can be solved by
methods of classical information theory, but in the second case new physical
representations are needed.\\ Mutual information is the most important
ingredient of information theory. In classical theory this value was
introduced by C.Shannon [9]. The mutual information between two ensembles of
random variables $X$, $Y$ (for example this ensembles can be input and
output for a noisy channel) 
\begin{equation}
I(X,Y)=H(Y)-H(Y/X),
\end{equation}
is the decrease of the entropy of $X$ due to the knowledge about $Y$, and
conversely with interchanging $X$ and $Y$. Here $H(Y)$ and $H(Y/X)$ are
Shannon entropy and mutual entropy [9].\\ Mutual information in the quantum
case must take into account the specific character of the quantum
information as it is described above. The first reasonable definition of
this quantity was introduced by B.Schumacher and M.P.Nielsen [2]. Suppose a
quantum system with density matrix 
\begin{equation}
\label{l1}\rho =\sum_ip_i|\psi _i\rangle \langle \psi _i|,\ \ \sum_ip_i=1.
\end{equation}
We only assume that $\langle \psi _i|\psi _i\rangle =1$ and the states may
be nonorthogonal. The noisy quantum channel can be described by a general
quantum evaluation operator $\hat S$ with kraussian representation 
\begin{equation}
\label{l2}\hat S\rho =\sum_\mu A_\mu ^{\dag }\rho A_\mu ,\ \ \sum_\mu A_\mu
A_\mu ^{\dag }=\hat 1.
\end{equation}
This operators must be linear, completely positive and trace-preserving
[1,4]. As follows from definition of quantum information transmission, a
possible distortion of entanglement of $\rho $ must be taken into account.
In other words definition of mutual quantum information must contain the
possible distortion of relative phases of quantum ensemble $\{|\psi
_i\rangle \}$. Mutual quantum information is defined as [2] 
\begin{equation}
I(\rho ;\hat S)=S(\hat S\rho )-S(\hat 1^R\otimes \hat S(|\psi ^R\rangle
\langle \psi ^R|)),
\end{equation}
\begin{equation}
\hat 1^R\otimes \hat S(|\psi ^R\rangle \langle \psi ^R|))=\sum_{i,j}\sqrt{%
p_ip_j}|\phi _i^R\rangle \langle \phi _j^R|\otimes \hat S(|\psi _i\rangle
\langle \psi _j|).
\end{equation}
Where $S(\rho )$ is the entropy of von Newman and $\psi ^R$ is a
purification of $\rho $ 
\begin{equation}
|\psi ^R\rangle =\sum_i\sqrt{p_i}|\psi _i\rangle \otimes |\phi _i^R\rangle
,\ \ \langle \phi _j^R|\phi _i^R\rangle =\delta _{ij},
\end{equation}
\begin{equation}
tr_R|\psi ^R\rangle \langle \psi ^R|=\rho ,
\end{equation}
here $\{|\phi _i^R\rangle \}$ is some orthonormal set. The definition is
independent from concrete choice of this set [1]. The mutual quantum
information is the decrease of the entropy after acting of $\hat S$ due to
the possible distortion of entanglement state. This quantity is not
symmetric with respect to interchanging of input and output and can be
positive, negative or zero in contrast with the Shannon mutual information
in classical theory.\\ It has been shown that (4) can be the upper bound of
the capacity of a quantum channel [3,11]. Using this value the authors [3]
have been proved the converse coding theorem for quantum source with respect
to the entanglement fidelity [1]. Only this fidelity is adequate for quantum
data transmission or compression.\\ In the [2] the authors prove data
processing inequality 
\begin{equation}
I(\rho ;\hat {S_1})\geq I(\rho ;\hat {S_2}\hat {S_1}).
\end{equation}
In [3] we found the alternative derivation of this result which is more
simple than derivation of [2]. In this paper we show that this equation can
be strengthened. Data processing inequality is very important property of
mutual information. This is an effective tool for proving general results
and the first step toward identification a physical quantity as mutual
information.\\ Now we brief recall the derivation of data processing
inequality in the general case. The formalism of relative quantum entropy is
very useful in this context [5,7].\\ Quantum relative entropy between two
density matrices $\rho _1$, $\rho _2$ is defined as follows 
\begin{equation}
\label{a1}S(\rho _1||\rho _2)=tr(\rho _1\log \rho _1-\rho _1\log \rho _2).
\end{equation}
This quantity was introduced by Umegaki [6] and characterizes the degree of
'closeness' of density matrices $\rho _1$, $\rho _2$. The properties of
quantum relative information were reviewed by M.Ohya [5]. Here are mentioned
only two basic properties 
\begin{equation}
S(\rho _1||\rho _2)\geq S(\hat S\rho _1||\hat S\rho _2).
\end{equation}
\begin{equation}
S(c\rho _1+(1-c)\sigma _1||c\rho _2+(1-c)\sigma _2)\leq cS(\rho _1||\sigma
_1)+(1-c)S(\rho _2||\sigma _2).
\end{equation}
Where $0\leq c\leq 1$. The first inequality was proved by Lindblad [7]. We
have 
\begin{eqnarray}
&   & S(\hat{1}^R\otimes \hat S(|\psi ^R\rangle \langle \psi ^R|)||\hat
{1}^R\otimes \hat S(\rho ^R\otimes \rho  ))\nonumber \\
& =&-S(\hat{1}^R \otimes \hat S(|\psi
^R\rangle \langle \psi ^R|))+S(\rho ^R)+S(\hat S\rho )).
\end{eqnarray}
Here 
\begin{equation}
\rho ^R=\sum_{i,j}\sqrt{p_ip_j}|\phi _i^R\rangle \langle \phi _j^R|\langle
\psi _i|\psi _j\rangle .
\end{equation}
Now from Lindblad inequality we have 
\begin{eqnarray}
&      & S(\hat {1^R}\otimes \hat S(|\psi ^R\rangle \langle \psi ^R|)||\hat
{1^R}\otimes \hat S(\rho ^R\otimes \rho ))\nonumber \\
& \geq & S(\hat {1^R}\otimes \hat
{S_1}\hat {S_2}(|\psi ^R\rangle \langle \psi ^R|)||\hat {1^R}\otimes \hat
{S_1}\hat {S_2}(\rho ^R\otimes \rho )).
\end{eqnarray}
From this formula we have (8). From (12) we have 
$$
{}I(\rho ;c\hat S_1+(1-c)\hat S_2)\leq cI(\rho ;\hat S_1)+(1-c)I(\rho ;\hat
S_2) 
$$
This theorem have been proved in [11].\\ Now we use some general theorems
for the strengthening the ordinary data processing inequality. In the first
we strengthen the Lindblad inequality.\\ Let us assume in formula (10) that 
\begin{equation}
\hat S=c\hat {C_1}+(1-c)\hat {C_2},
\end{equation}
where $\hat {C_1}$ is defined by kraussian representation $A_\mu =|\mu
\rangle \langle 0|,\ \ \langle \mu |\acute \mu \rangle =\delta _{\mu \acute
\mu },\ \ \langle 0|0\rangle =1,\ \ 0\leq c\leq 1$. In other words for any
operator $\rho $ $\hat {C_1}\rho =|0\rangle \langle 0|.$ Now from (10), (11)
we get 
\begin{eqnarray}
&     & S(\hat S\rho ||\hat S\sigma)=
S(c\hat {C_1}\rho +(1-c)\hat {C_2}\rho||c\hat {C_1}\sigma 
+(1-c)\hat {C_2}\sigma )\nonumber \\
&\leq & cS(\hat {C_1}\rho ||\hat {C_1}\sigma ) 
+ (1-c)S(\hat {C_2}\rho  || \hat {C_2}\sigma )
\leq (1-c)S(\rho ||\sigma).
\end{eqnarray}
We see that if $\hat S$ is represented in the form (15) the ordinary
Lindblad inequality can be strengthened.\\ Now we need some general results
from operator theory [10]. Let two hermitian operators $A$ and $B$ have the
spectrums $a_1\leq ...\leq a_n$, $b_1\leq ...\leq b_n$. For the spectrum $%
c_1\leq ...\leq c_n$ of the operator $C=A+B$ we have 
\begin{equation}
a_1+b_k\leq c_k\leq b_k+a_n,\ \ b_1+a_k\leq c_k\leq a_k+b_n.
\end{equation}
where $k=1,...,n$. If 
\begin{eqnarray}
&     &\rho \prime =\hat {S}\rho = c\hat {C_1}\rho +(1-c)\hat {C_2}\rho \nonumber \\
& =   & c| 0\rangle   \langle 0 |+(1-c)\sigma ,
\end{eqnarray}
and $\rho _1\prime \leq ...\leq \rho _n\prime $, $\sigma _1\leq ...\leq
\sigma _n$ are the spectrums of $\rho \prime $, $\sigma $ then we have 
\begin{eqnarray}
&   &\rho _{1}\prime -c\leq \sigma _{1}(1-c)\leq \min ( \rho _{1} \prime ,
\rho _{n} \prime -c),\nonumber \\
&   & \max (\rho _{1}\prime ,\rho _{k}\prime -c)\leq \sigma _{k}(1-c)\leq
\rho _{k} \prime ,
\end{eqnarray}
where $k=2,...,n$. We define $c(\hat S,\rho )$ as the minimal eigenvalue of $%
\rho \prime $ and $c(\hat S)=\min _\rho c(\hat S,\rho )$ where minimization
is taken by all density matrices for the fixed Hilbert space. With the well
known results of operator theory [10] we can write 
\begin{equation}
c(\hat S)=\min _\rho \min _{\langle \psi |\psi \rangle =1}\langle \psi |\hat
S\rho |\psi \rangle ,
\end{equation}
where the second minimization is taken by all normal vectors in the Hilbert
space. For any density matrix $\rho $ we get to the formula (15) 
where c is defined in (20) and $\hat {C_2}$ is some general evolution
operator. Now from (15, 16, 20) we get the strengthened Lindblad inequality 
\begin{equation}
(1-c)S(\rho _1||\rho _2)\geq S(\hat S\rho _1||\hat S\rho _2).
\end{equation}
Now we can prove the strengthened data processing inequality. Let in (14) $%
\hat {S_2}$ is represented in the form (15). From (10-15) we get 
\begin{eqnarray}
&   & S(\hat{1}^R\otimes \hat S_2\hat S_1(|\psi ^R\rangle \langle \psi ^R|)
||\hat {1}^R\otimes \hat S_2\hat S_1(\rho ^R\otimes \rho  ))\nonumber \\
& \leq &-(1-c)S(\hat{1}^R \otimes \hat C_2\hat S_1(|\psi
^R\rangle \langle \psi ^R|))+S(\rho ^R)+(1-c)S(\hat C_2\hat S\rho )).
\end{eqnarray}
And we have 
\begin{equation}
(1-c(\hat {S_2}))I(\rho ;\hat {S_1})\geq I(\rho ;\hat {S_2}\hat {S_1}).
\end{equation}
The equation (20), (23) are our final results. Of course there are many
evolution operators $\hat S$ with $c(\hat S)=0$ but now we show that our
results can be nontrivial because for some simple but physically important
case $c(\hat S)$ is nonzero.\\ Now we consider the simplest example of noisy
quantum channel: Two dimensional, two- Pauli channel [8] 
\begin{equation}
A_1=\sqrt{x}\hat 1,\ \ A_2=\sqrt{(1-x)/2}\sigma _1,\ \ A_3=-i\sqrt{(1-x)/2}%
\sigma _2,\ \ 0\leq x\leq 1,
\end{equation}
where $\hat 1$, $\sigma _1$, $\sigma _2$ are the unit matrix and the first
and the second Pauli matrices. Any density matrix in two-dimensional Hilbert
space can be represented in the Bloch form 
\begin{equation}
\rho =(1+\vec a\vec \sigma )/2,
\end{equation}
where $\vec a$ is a real vector with $|\vec a|\leq 1$. Now we have 
\begin{equation}
\hat {S_{TP}}((1+\vec a\vec \sigma )/2)=(1+\vec b\vec \sigma )/2,
\end{equation}
where $\vec b=(a_1x,a_2x,a_3(2x-1))$. After simple calculations we get 
\begin{equation}
c(\hat {S_{TP}})=(1-|2x-1|)/2.
\end{equation}
We conclude by reiterating of the main result: The quantum data processing
inequality can be strengthened. Such strengthened inequality also exist in
classical information theory [9].

\end{document}